# Control of single spin in CMOS devices and its application for quantum bits.

**R. Maurand, D. Kotekar-Patil, A. Corna, H. Bohuslavskyi, A. Crippa, R. Laviéville, L. Hutin, S. Barraud, M. Vinet, S. De Franceschi , X. Jehl, M. Sanquer**

**Univ. Grenoble Alpes, INAC-PHELIQS, F-38000 Grenoble, France**
**CEA, INAC-PHELIQS, F-38000 Grenoble, France**
**CEA, LETI MINATEC campus, F-38000 Grenoble, France**

We show how to measure and manipulate a single spin in a CMOS device fabricated in a pre-industrial 300 mm CMOS foundry. The device can be used as a spin quantum bit working at very low temperature. The spin manipulation is done by a microwave electric field applied directly on a gate.
The presented results are a proof-of-principle demonstration of the possibility to define qubits by means of conventional industrial fabrication processes.

## 1. Introduction

Since the pioneering work of Scott-Thomas et al. [Scott89] it has been observed that carriers can be manipulated one by one in the channel of silicon field effect devices. However that takes a long time to be achieved in a controlled way in CMOS devices. The first time it has been realized on purpose in silicon-on-insulator channel was in ref. [Ali94] and in ref. [Takahashi95]. The first time it has been obtained in a CMOS transistor was in ref. [Boeuf03]. Since that time -at Grenoble- we have optimized a way to change a standard CMOS field effect transistor into a MOS-Single Electron Transistor (MOS-SET). The MOS-SET inherits from the main figures of merit of its companion device, i.e. excellent electrostatic control by the gate voltage, compactness, standardization, excellent variability and yield, fast and energy efficient operation [Hofheinz06] (see figure 1). This technique made the MOS-SET an ideal platform for analogue applications, as the silicon electron pump [Jehl13] or hybrid SET-FET amplifiers [Lavieville16] amongst many others [Takahashi02, Gautier09]. Already we have shown that a classical CMOS analogue electronics can be fully co-integrated with a MOS-SET [Clapera15].

Nevertheless the most radical application of the MOS-SET is probably its use as the elementary brick to build an all-silicon quantum computer.

One of the earliest -and today most studied- proposals for quantum computation in semiconductors envisioned arrays of electrostatically defined dots, each containing a single electron whose two spin states provide a qubit [Loss98]. Quantum logic is accomplished by changing voltages on the electrostatic gates to move electrons closer and further from each other, activating and deactivating the exchange interaction. Most of the spin qubit studied are focused on qubits in both GaAs/AlGaAs [Shulman12] and Si/SiGe heterostructures [Kawakami14] embedding a buried two-dimensional electron gas whose properties are tailored by band diagram engineering. These structures can be fabricated using lab-scale lithography tools. From large-scale integration perspective, however, III-V materials are not yet an option and the use of SiGe demands an adaptation with respect to conventional CMOS processes. Fortunately for this large-scale integration perspective a very large coherence time for the spin of electrons trapped on quantum dots in silicon, -which can be isotopically purified [Itoh14]- have been recently demonstrated [Veldshorst14]. As envisaged in the Loss–DiVincenzo proposal [Loss98], two qubit gates for electron spins in isotopically purified silicon quantum dots have been further realized [Veldshorst15].

Despite these fantastic advances, the silicon qubit is still a challenger in the race towards quantum computing [Ladd10] compared to superconducting solid state qubits [Lucero12] or trapped-ion qubits [Lanyon11]. Nevertheless silicon qubits present important advantages in terms of size ( a few 10 nm instead of microns), scalability and co-integration. This is particularly true if the silicon qubit can be realized in a full CMOS line.

This is the goal of this chapter to describe first how we control single spin in CMOS devices and second how we use it as a quantum bit. This is not the first qubit nor the most efficient, but -as already stated- when it comes to a crucial issue such as large-scale integration, however, the range of possible choices for the qubit becomes much narrower and the CMOS spin qubit becomes a serious option. The quantum computer is certainly a long term goal but the mere fact that a single spin can be controlled, manipulated and read out in a MOSFET device fabricated in the same foundries used for standard microelectronics is remarkable and was not anticipated at all only few years ago.

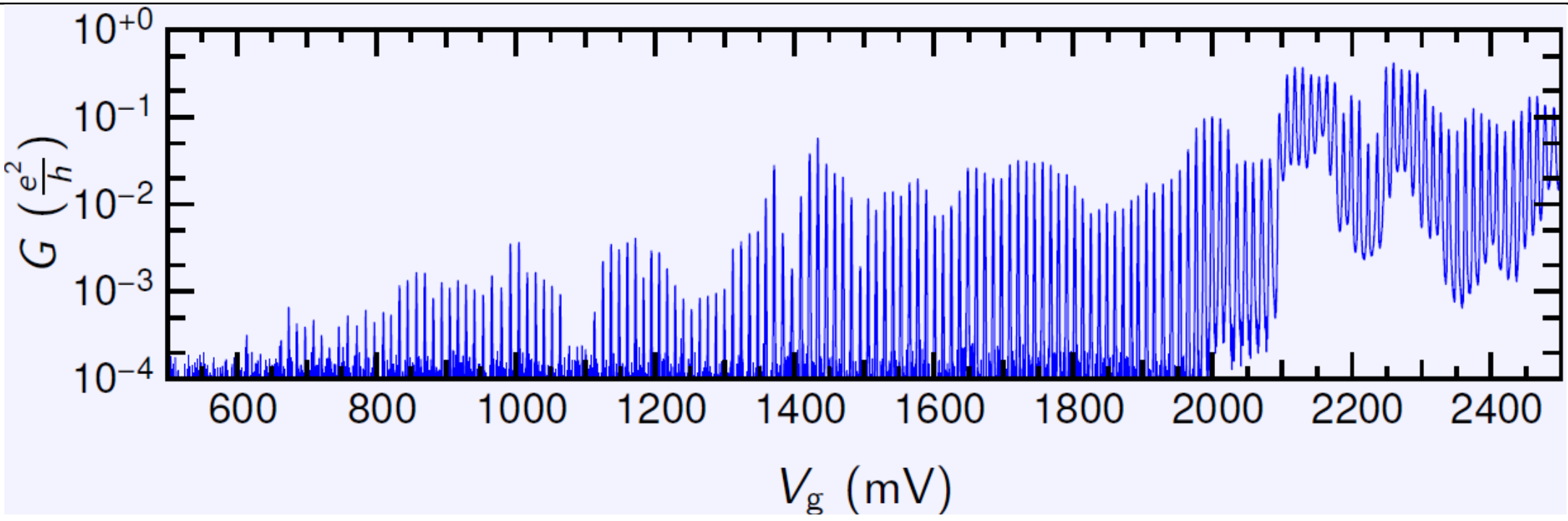

Figure 1: Periodic Coulomb oscillations of the drain-source conductance

(in quantum units $e^2/h \approx 1/25800\ \Omega$) versus gate voltage observed in a MOS-FET with large nitride spacers (50nm thick, non-overlapped geometry [Boeuf03]) at T=60mK. The drain voltage Vd is small enough to be in the linear Id-Vd regime. The length of the channel is 30nm and the width of the channel is 40 nm. The MOS-FET is done at the CEA-LETI using 24 nm thick silicon-on-insulator nanowire trigate technology. More than 200 oscillations can be recorded corresponding to the addition of electrons one-by-one in the channel from zero to a density of few $10^{13}$ carriers/cm$^2$ ($1.6\ 10^{13}$ e/cm$^2$ $\approx$ 200e/ (30 nm$\times$ 40nm)). The period in Vg is given by $\Delta Vg = e/Cg$ where Cg is approx. given by the planar gate-channel capacitance (sample similar to the one presented in ref. [Hofheinz06]).

## 2. Control of single spin in CMOS devices

The charge of a single electron has been measured for long time [Millikan1911]. But the tiny spin of a single carrier corresponds to an extremely small magnetic moment whose magnetization cannot be measured directly. It is necessary to perform first a spin-to-charge conversion. In the context of electronic devices this is done at very low temperature either by using a spin selective or an energy selective tunnelling event. In our measurement we used the spin selective scheme known as the Pauli blockade [Tarucha02]. The energy selective scheme, known as the Elzerman protocol [Elzerman04] relies on the Zeeman energy difference between an electron with the spin up or down. It is used for instance in the experiments by Veldhorst et al. [Veldshorst14, Veldshorst15].

The Pauli blockade prevents the tunnelling of one electron between two dots corresponding to the transition T(1,1)➔T(0,2) when the Triplet T(0,2) is too high in energy ((n,m) denotes the number of carriers in (dot1,dot2)) [Tarucha02]. The T(0,2) state is above the S(0,2) state because of the Pauli exclusion principle: to have two parallel spins, the electrons should occupy two different orbital states, the ground state and an excited state, that costs an additional kinetic energy term (minored by the possible exchange energy). In our dots this energy difference is of the order of one meV (about $k_B \times 12K$), because our dots are extremely small. This Pauli blockade scenario is well established when there is no or weak spin-orbit interaction [Tarucha02], that is true for electrons in silicon. In that case during the tunnelling event from one dot to the next the spin of the electron is conserved.

If there is a significant spin-orbit interaction –as it is the case for holes in silicon- the spin is possibly not conserved during the tunnelling but a reminiscent Pauli blockade persists [Li16, Bohuslavskyi16]. This spin blockade is partly dependent on the magnetic field. Spin blockade is stronger at B = 0 compared to finite magnetic field due to time-reversal symmetry. As a result, a current dip at B = 0 is expected [Bohuslavskyi16, Li16]. This is in contrast with the case of small spin-orbit coupling where the spin blockade can be partially removed at B=0 due to hyperfine [Koppens05] or cotunneling mechanisms [Qassemi09].

The Pauli blockade is used to initialize and read out the spin located in one of the two quantum dots. This is illustrated on fig. 2. <u>Initialization:</u> the gate 1 (resp. gate 2) -controlling the number of carriers in the dot 1 (resp. 2)- are polarized such that a carrier (a hole in the present experiment [Maurand16]) is sitting under gate 2 and the two (0,1) states, the four (1,1) states and S(0,2) state are energy degenerate. Without spin consideration there will be a finite drain-source current at these gate voltages through the (0,1)➔(1,1)➔(0,2)➔(0,1) sequence (at finite drain –source voltage). S(1,1)=T(1,1) are almost degenerate because the exchange coupling between the two dots is small. In the presence of a static magnetic field the spin on dot 2 is in the ground state –say spin down. As long as a spin up enters in the dot 1 it is transmitted into the drain. As soon as a spin down enters, the transfer is blocked. After a certain time of initialization, we can therefore be certain that the initial state is $T^-(1, 1)$, i.e. both spins down. This sequence corresponds to the two first steps of Fig 2 (upper part).

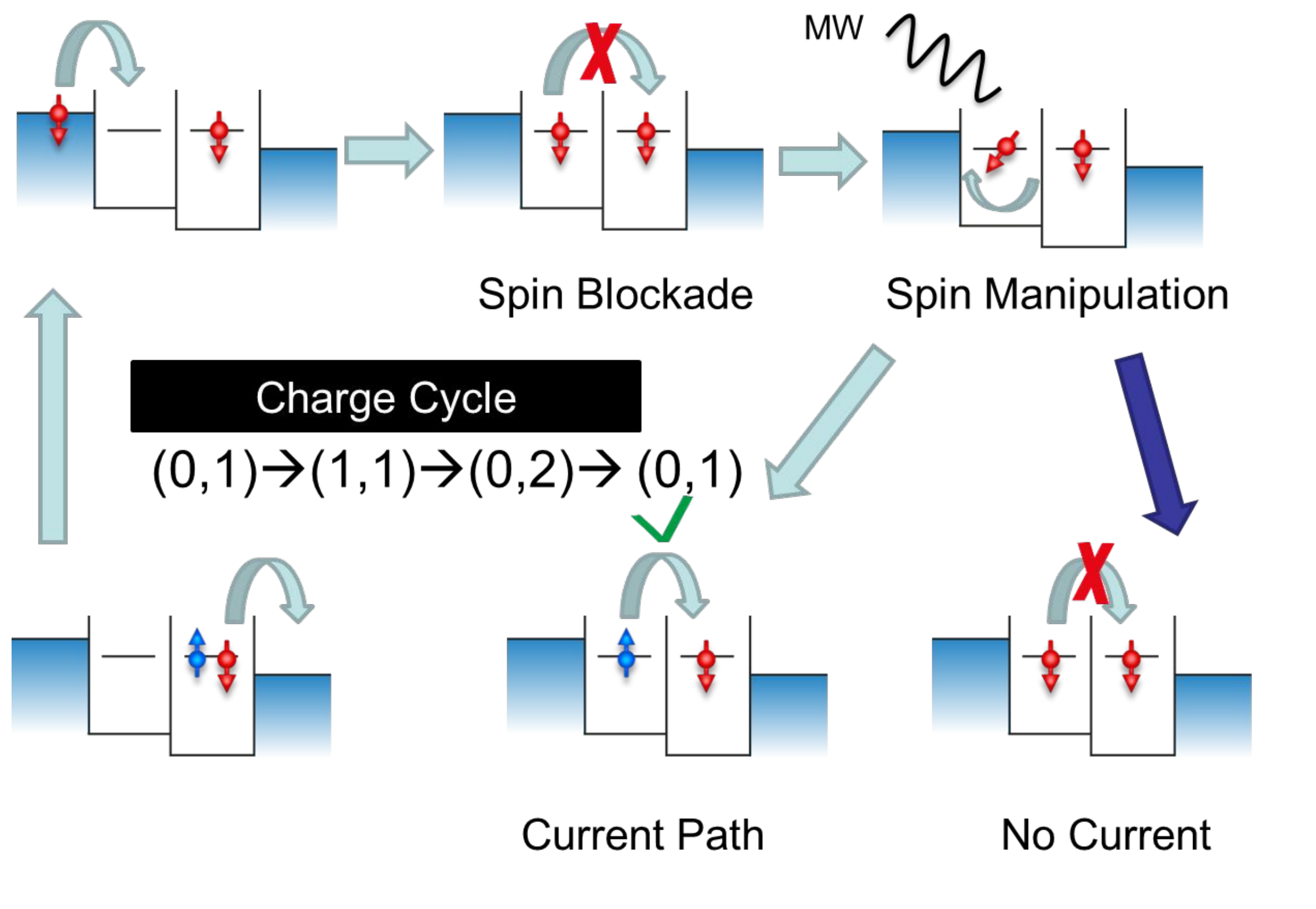

Figure 2: Principle of the Pauli spin blockade in two quantum dots in series and the principle for ESR or EDSR detection by the source-drain current. The diagram is made for electrons. <u>Top-left</u>: (0,1)➔(1,1) transition; the current is blocked as soon as a carrier with a spin down enters in dot 1 (top-center). <u>Top right:</u> thanks to an electric (resp. magnetic) radiofrequency voltage (at the right frequency, $h=g❑_B H_0$ ) applied on gate 1, the spin rotates. For an RF

signal applied *continuously* the spin is up after some time and a DC drain source current due to EDSR (resp. ESR) is detected (bottom center and left, the indicated charge cycle with light blue arrows is continuously working). Bottom right (dark blue arrow): the radiofrequency voltage is applied during a *fixed* time $\tau$ (RF Burst). If $\tau$ corresponds to a $2N\pi$ rotation the spin is still down at the end of the burst and there is no current. To avoid a transient leakage during the burst and/or to detect several rotations (see fig. 6) the ESR or EDSR burst is applied when the two carriers are in the Coulomb blockade regime (the spin manipulation step is done when two levels below the Fermi energy in the drain (not shown), as the initialization and readout are done in the Pauli blockade regime, see section 3)

The Pauli blockade is also used for Readout:

First the (magnetic) Electron spin resonance (ESR) or the electric dipole spin resonance (EDSR) is detected when a continuous microwave signal at the right frequency $h = g\mu_B H_0$ is applied on gate1. This corresponds to the cycle shown by the light blue arrows in Fig. 2. Second the Rabi oscillations can be recorded if a microwave burst is applied when the two dots are in the Coulomb Blockade regime as explained in the next section and featured by the dark blue arrow. Pauli spin blockade then offers a way to converse the information on a single spin orientation on dot 1 into a charge transfer event, possibly a measurable current after the integration of many such events.

Note that the Pauli blockade is efficient as long as the temperature is less than the energy separation between S(0,2) and T(0,2) (≈1meV). On the contrary the Elzerman protocol necessitates that $k_BT<g\mu_BH$ and the Zeeman splitting is much smaller than the singlet-triplet energy separation in general. With the Pauli blockade scheme it is possible to detect the spin at a temperature of about 1K as for the Elzerman protocol [Elzerman04] one should go below T=0.1K. This could make a significant difference in the power consumption budget to have qubits and their peripherals working at T=1K instead of 0.1K.

With the Pauli blockade it is possible to initialize and read out a single hole or electron spin and with ESR or EDSR it is possible to manipulate the spin. To be useful for applications the spin orientation should be stable enough between successive manipulations. Electron spin in silicon are remarkably insensitive to their electrostatic environment that makes them very interesting as quantum bits. This has been observed for long time in macroscopic silicon crystals doped with donors and studied by Electron Spin Resonance (ESR) standard techniques (where a huge number of spins are manipulated coherently). The spin relaxation time $T_2$ can reach 0.4s in very diluted Si:P at 1.2 $10^{14}$ $cm^{-3}$ if the silicon crystal is purified from the $^{29}Si$ isotopes [Witzel10]. Nuclear spin relaxation time can be much longer: $T_{2N}$=192s at T=1,8K in $^{28}Si$:P (5 $10^{11}cm^{-3}$) [Steger12] . Therefore isotopically purified $^{28}Si$ crystal can be considered as a “silicon vacuum” for electron spins and an excellent platform for spin qubits.

In macroscopic silicon crystals the electron are localized on donors. In nanoscopic transistors the carriers can be either localized on quantum dots or donors. Remarkably for electrons in silicon quantum dots, $T_2$ reaches 28 ms [Veldshorst14]. Even is this time is smaller (by a factor of 10) compared to a macroscopic lightly n-doped $^{28}$Si crystal, it is remarkably long if one considers that electron lives in an artificial nanostructure with nearby gates, electrodes, interfaces, defects, etc. The $T_2$ and the inhomogeneous $T_2^*$ relaxation times are typically $10^3$ times shorter in natural silicon, pointing the importance of the hyperfine coupling with nuclear spins to explain the electron spin relaxation time limitation at low temperature in natural silicon nanostructures. These remarkably long spin coherence times observed for electrons in silicon nanostructures pushed us to look after spin quantum bits in CMOS nanostructures fabricated in a pre-industrial platform devoted to classical nano-electronics.

We concentrate on carriers confined in quantum dots rather than on dopants because it is difficult to control a single dopant with enough precision. This is possible by STM assisted nano-patterning [O'Brien01] but this elegant technique is not scalable and hardly compatible with standard CMOS techniques, that is our main objective. We notice nevertheless that standard CMOS techniques associated with controlled channel doping permitted to build single–atom and coupled-atom transistors [Zwanenburg13,Sellier06,Pierre10,Roche12]. Therefore it would be possible to use dopants to develop new functionalities for the silicon qubit, for instance to store quantum information on nuclear spins which have a much longer coherence time than electron spins [Muhonen14, Morton08].

Furthermore we choose the Fully-Depleted Silicon-on-Insulator (FDSOI) technology, and its variant -the trigate technology- to build our qubits. This is illustrated on figure 3.

The silicon nanowire field-effect transistors (NW-FETs) are fabricated on a 300mm Silicon-On-Insulator (SOI) processing line [Barraud2012]. First, a silicon nanowire is etched from a SOI wafer with a 10-nm-thick, undoped silicon device layer. The nanowire channel is oriented along the [110] direction. Initially defined by deep ultra-violet (DUV) lithography, its width W is trimmed down to about 15 nm by a controlled oxidation and etching process. Two parallel top-gates, ≈35-nm wide and with a ≈30 nm spacing between them, are successively patterned by means of a combined DUV and e-beam lithography. The latter enables us to achieve the necessary small spacing between the gates. The gate stack consists of a thin (≈5 nm) TiN layer followed by a much thicker (≈50 nm) polysilicon layer. Gate electrical isolation is ensured by a dielectric stack consisting of a $SiO_2$ layer of 7 nm and an Hf-based high-$\kappa$ dielectric layer of 2 nm. Insulating SiN spacers are deposited all around the gates. Their width is deliberately large in order to fully cover the nanowire channel between the two gates and protect it from the successive ion implantation process, which is required for low resistance ohmic contacts to the nanowire channel. For these p-type devices we use boron ion implantation. Wide spacers also limit boron diffusion from the heavily implanted contact regions into the channel. Dopants are activated by spike annealing followed by self-aligned silicidation. Devices are finalized with a standard microelectronics back-end of line process. At the end, the whole device fabrication is based on standard processes of our CMOS line, except for the e-beam lithography. We note that gate pitches as small as the one used here, i.e. well below the diffraction limit of DUV (about 190 nm), could as well be obtained with DUV through multiple patterning combined with high-precision realignment [Natarajan 14]. A schematic representation of the encapsulated device is shown on fig 3.

This trigate technology presents several decisive advantages:

First, this technology has been introduced to preserve a perfect control of the electrostatic potential in the silicon channel below the gate, even at small gate length. This control is decisive also in our spin qubits because we manipulate the spin with voltages applied on the gates (see later on). The more perfect is the control of the quantum dot potential with the gate, the lower is the signal strength to control the spin. Moreover the excellent “electrostatic integrity” means that the potential of the dot is less sensitive to the potential applied on nearby gates that is crucial for upscaling the qubit. It is important that the signal applied on gate 1 does not influence too much the spin located below gate 2. Moreover we plan to measure the qubit by dispersive gate reflectometry in the future [Gonzalez16] (see section 4) and this technique relies on the exquisite coupling between the gate and the qubit.

Second, this technology allows us to use the substrate bias to control the electrostatic potential in the nanowire in three dimensions: By applying a positive substrate bias (across the buried oxide (BOX)) we can for instance push the holes near the top of the nanowires where the first holes appear in the corners of the nanowire below the front gate, see section 4 [Voisin15]. This can be used to create two quantum dots in parallel along the channel (see fig. 8) which can be used either to encode a qubit (single-triplet qubit) or to detect the spin orientation using Pauli blockade and gate reflectometry [Gonzalez16].

Third, in the case of two gates in series along the nanowire channel, the substrate bias is used to control the tunnelling rate between the two dots (and between the dots and the source/drain) by modulating the potential in the part of the nanowire which are not covered by the front gates. These tunnelling rates should be adjusted such that the carriers can be transmitted between the source, the two dots and the drain at a sufficiently high rate to detect the source-drain current and at sufficiently low rate to localize the wave functions below the each gate (as far as we consider the single spin qubit and not a singlet-triplet coupled dot qubit with adjustable exchange term, of course).

The trigate nanowire technology permits to manipulate either electrons or holes, depending on the nature of the source-drain doping (n or p-type) and on the polarity of the gate voltage. For our qubit we decided to use holes instead of electrons. Up to now there is no report of an hole qubit whatever the semiconductor material. Continuous EDSR of holes has been observed in [Pribiag13]. The choice for holes is then rather extreme but motivated by a decisive advantage that an hole spin can be manipulated by an electric field as electron spin cannot in principle. Because in CMOS devices the use of magnetic fields and magnetic coupling is very unusual (excepted for MRAM) as the electric field manipulation is the common rule this makes the hole spin qubit much more compatible with the standard CMOS than the electron spin qubit.

The penalty to use holes rather than electrons in silicon comes from the larger sensitivity of holes to static disorder (the lower hole mobility compared to electron mobility is the signature of its effect). We noticed that holes are more prone to be localized by residual disorder at the top of the valence band compared to electrons at the bottom of the conduction band. This is probably due to the larger hole effective mass but possibly also to the nature and charge states of the defects in the gate stack or in the spacers. As a consequence it is relatively easy to control the first electron in the MOS-SET, as it is much more difficult to certify that the first hole in a MOS-SHT (Single Hole Transistor) is detected through transport measurement: the first hole could be localized in a shallow asperity of the electrostatic confinement potential and too weakly coupled to the electrodes for detecting any current.

Compared to electrons, holes present both a much stronger spin-orbit coupling and an absence of valley-orbit coupling. Both features are important for the spin qubit. For a spin quantum bit it is important that orbitals are non-degenerate (except the two fold spin degeneracy of course) because intra orbital transitions can spoil the qubit integrity. For electrons at the $Si/SiO_2$ interface in the presence of a vertical electric field, the ground state orbitals is doubly degenerate (without counting the spin degeneracy, four times degenerate in total). Fortunately for the electron spin qubit this degeneracy is lifted by 3D electric field components in quantum dots or around a dopant. The typical energy spacing between the orbitals is 0.1-1meV, but very sensitive to the exact -and sometimes uncontrolled - microscopic configuration.

For holes in bulk silicon nevertheless the valence band is doubly-degenerate (heavy and light holes) [feher63] and this degeneracy is lifted by stress or electric field. In two-dimensional planar structures the ground state is spin-3/2 heavy-hole-like for instance [Winkler 03]. The minimal hypothesis in our structures is that the ground state is a Kramer's doublet with a mixing of spin-3/2 heavy and spin-1/2 light holes character.

The p-type character of holes in the valence band is responsible for the large spin-orbit coupling but also implies a reduced hyperfine coupling [Testelin09]. This makes the hole spin less sensitive to the nuclear spin in the silicon channel than the electron spin is. Therefore the isotopic purification of the silicon crystal can have less influence for hole spin coherence than it has for electron spin qubits, where it increases the spin decoherence time by a factor larger than $10^3$. Nevertheless the silicon isotopic effect for hole spins localized on boron acceptors in silicon has been studied in ref. [Stegner12]: local fluctuations of the valence-band edge due to different isotopic configurations in the vicinity of the boron acceptors account for inhomogeneous broadening effects of the ESR line in natural silicon crystals. Therefore hole spin resonance measurements in isotopically purified silicon quantum dot are very desirable in the future.

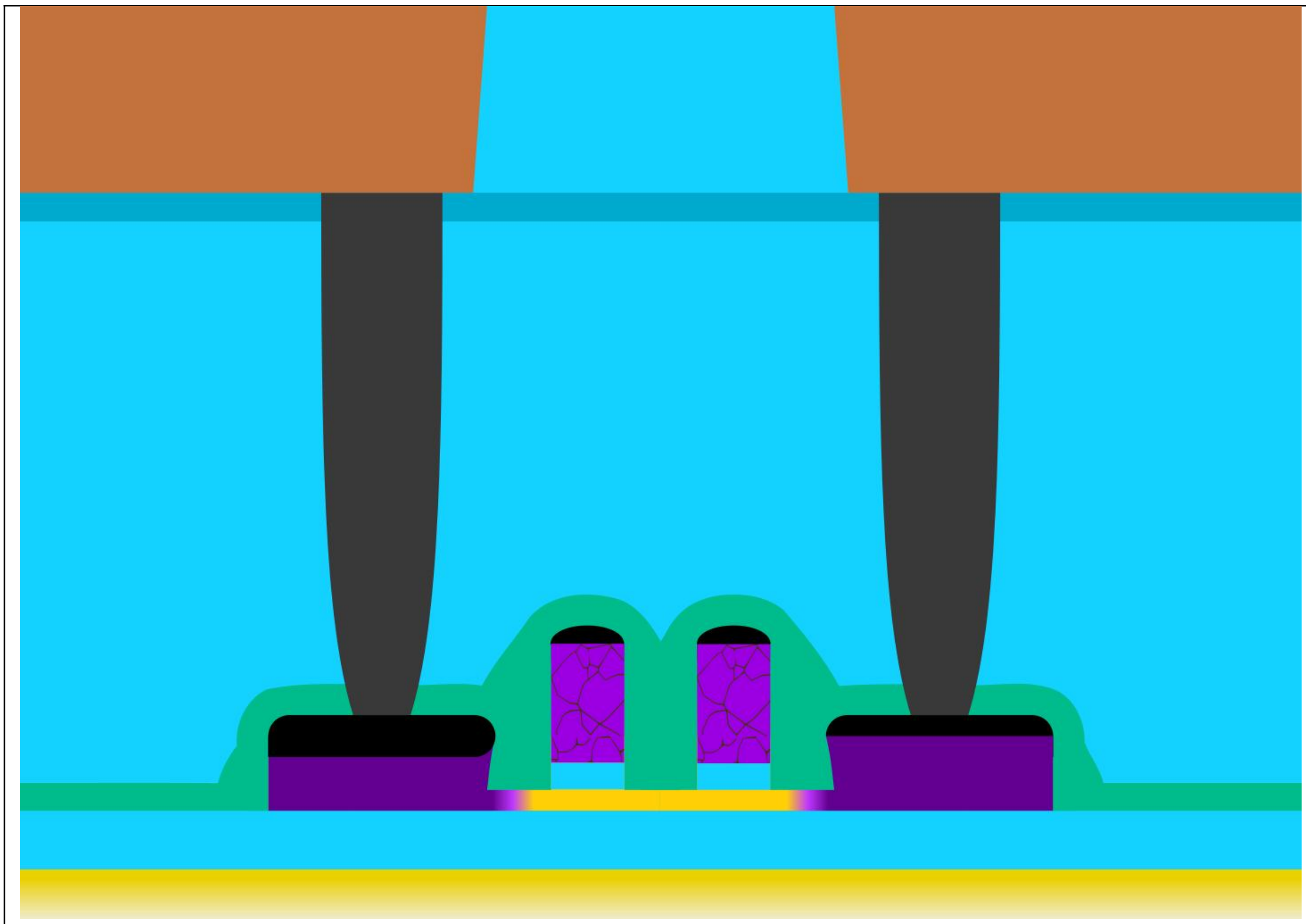

Figure 3: Schematic view of the coupled MOS-SET used for the CMOS hole spin quantum bit experiment. The thin silicon channel (≈10 nm, yellow) is covered by two top gates ( gate stack: $SiO_2$ in blue + thin (≈5 nm) TiN layer (not shown) + thick (≈50 nm) polysilicon layer in violet + silicide contact in black). The distance between the two gates is 35 nm. The re-grown source drain are heavily doped (violet ), silicided ( black) and contacted with metallic VIA's (black) to the metal 1 layer (orange). The oversized nitride spacers are featured in green. The 145 nm thick ( not to scale) buried $SiO_2$ oxide and the encapsulation oxide are featured in blue. The silicon substrate -on which the substrate bias is applied -is represented in yellow.

## 3. Hole spin qubit in CMOS devices

The main advantage of the hole spin is that it can be manipulated by an electric field at a fast rate. The hole spin is coupled with the electric field through spin-orbit coupling (even in the bulk case):

$$Hso = \frac{h}{4 m^2 c^2} \vec{Ez} (\dot{z} \times p)$$

It makes possible to change the spin orientation with an electric field and to perform Electric-Dipole Spin Resonance (EDSR). A more specific EDSR effect can exist in our nanowire PMOS, called the g-tensor modulation resonance (g-TMR). It corresponds to the case where the Landé g-factor is both anisotropic and varies with the electric field [Kato03]. Both the anisotropy and the voltage gate dependence of the g-factor have been observed in PMOS nanowire [Voisin16] and a Rabi frequency up to 500MHz has been predicted in that case. This Rabi frequency for EDSR is at least ten to hundred times larger than the Rabi frequency obtained for Electron Spin Resonance (ESR), about 100kHz [Veldhorst14]. This is a decisive advantage for the spin qubit where one should have large $T_2$ / $T_{Rabi}$ ratio, that controls the number of operations that can be performed on the spin qubit without losing quantum coherence by spin relaxation ($T_2$). Also $T_{Rabi}$ should be not too small compared to classical calculators (typically working at 1GHz) to make the quantum computers attractive enough.

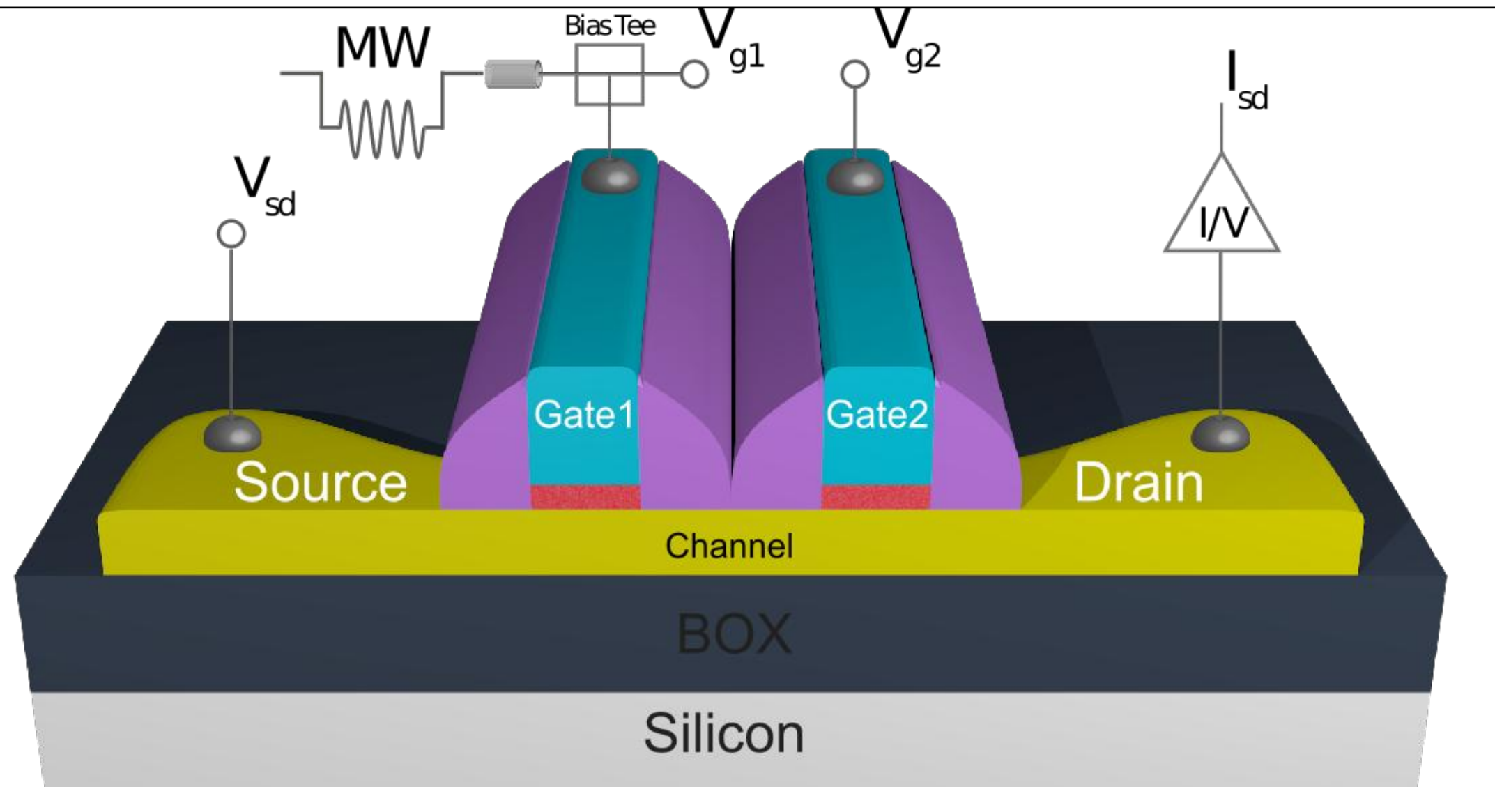

Figure 4: Schematic view of the electrical connections towards the coupled MOS-SET to control and read-out the hole spin qubit under the gate 1. The gate 2 is used to induce Pauli blockade (Spin selective tunnelling from dot 1 to dot 2). The read-out is done by the measurement of the DC drain-source current when a repetitive series of initialisation/manipulation/read out sequence is performed on gate 1 and 2 (repetition rate about 2MHz). From [Maurand2016].

The g-tensor modulation resonance (g-TMR) can be qualitatively explained as follows with a waving hand argument: in standard ESR one applies a RF magnetic field $H_{RF}$ perpendicular to a static magnetic field $H_0$. The ESR happens when $h=g\mu_B H_0$ where is the RF frequency and $\mu_B$ the Bohr magneton. For EDSR- in the presence of an RF *electric* field $\delta E_{RF}$ - the spin "sees" an equivalent *magnetic* field $H_{RF}$ given by $g\ H_{RF} = (\delta g / \delta E) \times \delta E_{RF}\ H_0$ (and because of the g-factor anisotropy, $H_{RF}$ has a component perpendicular to $H_0$).

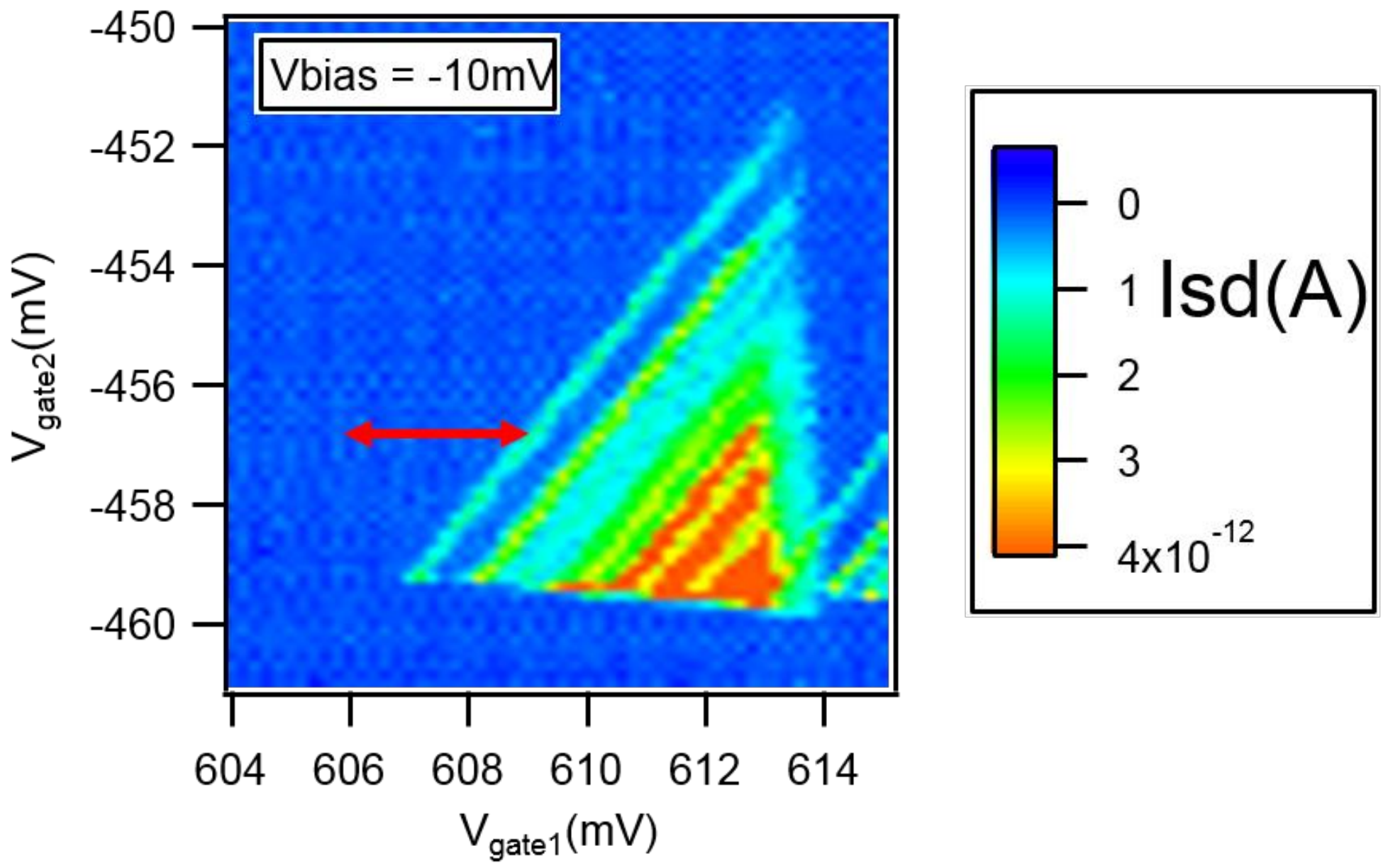

Figure 5: Drain-source current versus Vg1 and Vg2 at T=10mK and in absence of microwave on gate 1 (Vds=10mV). The "triangular region" defines the region where the two last empty levels in the dots are between the Fermi energy in the source and drain for the occupation numbers (0,1), (1,1) (0,2) ( see fig 2). The base of the triangle (along the diagonal) correspond to the alignment of the two ground levels in the dot, as the tip of the triangle (right lower corner) correspond to the maximum allowed detuning between these states: the ground state of the dot 1 aligned with the Fermi energy in the source and the ground state of the dot 2 aligned with the Fermi energy in the drain. The lines parallel to the base indicate when the ground state of the dot 1 is aligned with an excited state of the dot 2. Between these lines there is current only if inelastic processes between the ground state of dot 1 and one state of dot 2 are permitted (by the concomitant emission of photons /phonons). The small extra figure at the right lower corner is a replica of the main triangle due to the presence of an offset charge. The tips of the red arrow indicate the points where -by changing Vg1- we shift from the Coulomb blockade regime -i.e. (1,1) state below the Fermi energies both in source and drain ( left tip) -to the Pauli blockade regime - where the triplet T(1,1) is between the two Fermi energies, but

T(0,2) is not - (right tip).

The measurement scheme for the spin qubit is described in fig. 2 and fig. 4. It consists in measuring the source-drain current $I_{SD}$ for a given $V_{SD}$ in the few mV range at very low temperature (T < 0.1K), with DC and microwave gate voltages applied in a several sequences:

First DC voltage are applied to gate 1 and to gate 2 in such a way that the two last empty levels in the dots are between the Fermi energy in the source and drain for the occupation numbers (0,1), (1,1) (0,2) ( see fig 2). With *hole* quantum dot we cannot say that the occupation numbers are absolute numbers and –in fact- “0 holes” means 2N holes with N≈10-20. This distinction is not important as far as the orbital level spacing in the dot is much larger than the temperature and the Zeeman energy. This is realized in our hole quantum dots thanks to their very small size and their small density of states (compared to metals). These conditions (states (0,1), (1,1) (0,2) lying in the Fermi energy window) are realized in a “triangular region” in (Vg1,Vg2) plot at finite Vds (see fig. 5). The lines running parallel to the base of the triangle (along the diagonal on fig. 5) indicate when the ground state of the dot 1 is aligned with an excited state of the dot 2: the spacing between the ground state and the excited states of dot 2 is clearly resolved.Then we fix Vg2 and varies Vg1 as indicated by the red arrow on fig. 5. For Vg1≈ 606mV we are in the Coulomb blockade regime where the (1,1) state lies below the Fermi energies both in source and drain, as for Vg1≈ 609mV we are in the Pauli blockade regime - where the triplet T(1,1) lies between the two Fermi energies, but T(0,2) is not. We apply a static magnetic field of about 0.144T.

The sequence on Vg1 is the following: in the initialization phase we put Vg1≈ 609mV and we wait enough time (≈150 ns) such that we can be certain that the initial state is $T^{-}(1,1)$, as explained in section 2 ( fig. 2, top center panel). Then we put Vg1≈ 606 mV to be in the Coulomb blockade regime. We apply a microwave signal on Vg 1 during $\tau_{burst}$ (fig 2: spin manipulation top right panel done in the Coulomb blockade regime). The burst can be applied at the beginning or at the end of the Coulomb blockade sequence (duration 175 ns) without observed difference, indicating that the inelastic spin scattering time $T_1$ ( for flipping the spin orientation along the static magnetic field) is much longer than 175 ns. The mixing with the DC signal is done using a bias tee represented on fig. 4. The microwave frequency is varied across the resonant frequency for the EDSR $h = g\Box_B H_0$i.e. 8.938GHz. During the RF burst the spin of hole located under gate 1 is rotating on the Bloch sphere. Then we return after a time ≈175 ns (much larger than $\tau_{burst}$) on the Pauli blockade regime, Vg1≈ 609mV (fig. 2, bottom panels). Depending on the respective spin orientation for the hole on dot 1 and on dot 2 either a hole is transferred or not: if the state after the burst is T(1,1) there is no hole transfer, but if the state is S(1,1) one carrier is transmitted into the drain. For $\tau_{burst}$ ≈ 0, there is no transfer because the initial state is T(1,1). We choose a read-out time about 150ns such that the full sequence (initialization, manipulation, readout) lasts ≈435 ns. This fixes a limit to our detection because one hole transferred from source to drain each 450 ns correspond to a DC current of about 3.6pA. Too long time for the full sequence will result in poor signal-to-noise ratio, as shorter sequence will induce incomplete initialization or readout.

The integrated DC current (integration time = 1 s) as function of the RF burst duration is plotted on fig. 6.

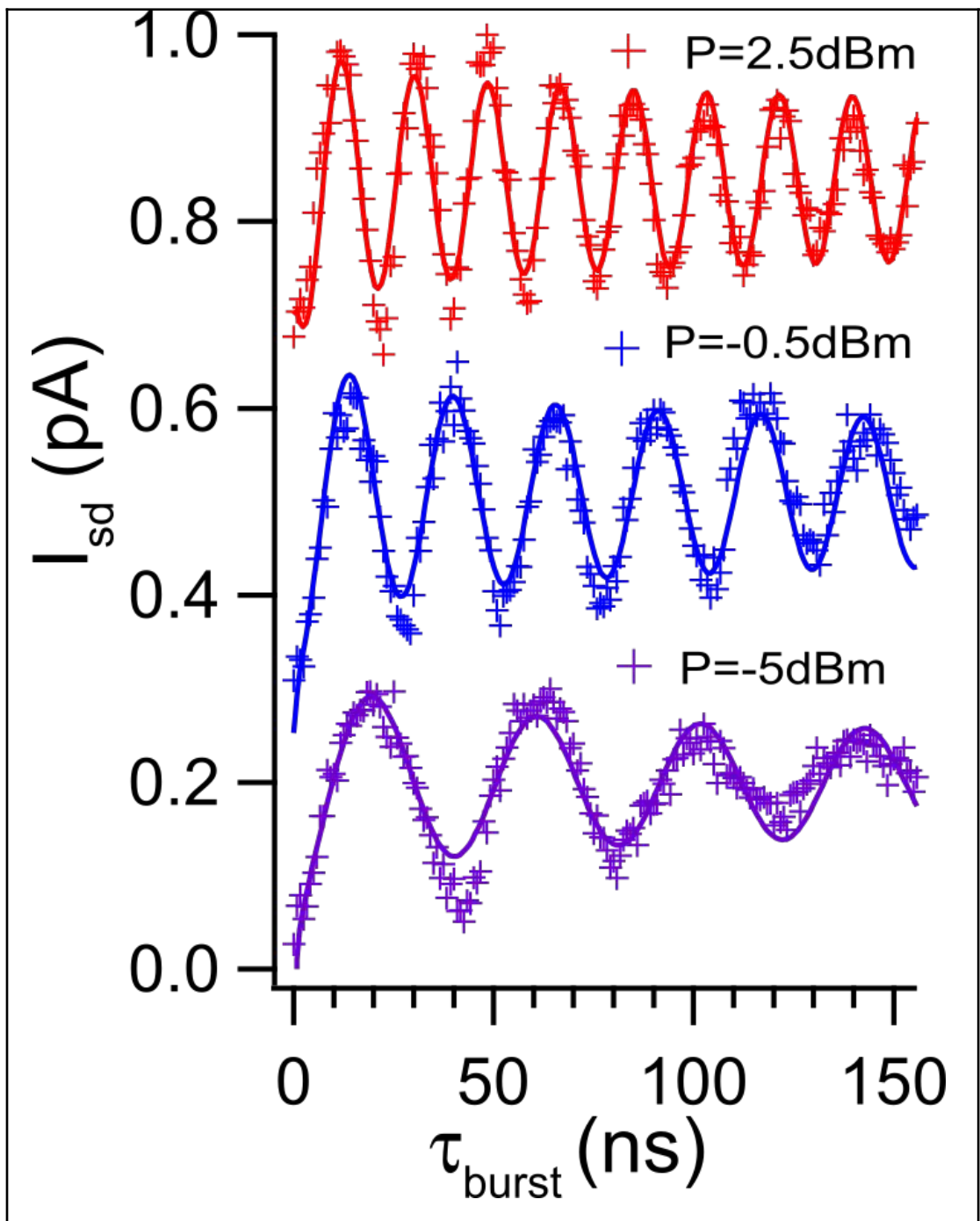

Figure 6 : Rabi oscillation of the hole spin qubit under gate 1, as a function of the burst time during which an RF signal (8.938GHz, H=0.144 T) is applied on gate 1, the coupled MOS-SET being in the Coulomb blockade regime.

The solid lines are fits as in [Koppens07]. Three powers for the RF signal are shown. As expected the Rabi frequency increases with the power, reaching 55 MHz for the largest applied power (2.5 dBm). Rabi frequencies are 24, 39 and 55MHz for $P_{MW}$=-5, -0.5 and 2.5 dBm, respectively.

From [Maurand2016].

As expected the detected current oscillates as function of $\tau_{burst}$. The inverse pseudo-period of the oscillations is the Rabi frequency. It increases linearly with the MW voltage amplitude (the square root of the voltage power). For our largest power it reaches 55MHz (and even 85MHz, not represented here [Maurand16]), that is much larger than for magnetic manipulation (ESR) of spin qubit in silicon (≈100 kHz, [Veldshorst14]).

Various spin relaxation times can be evaluated by making echo type of experiments. The inhomogeneous dephasing time $T_2$* can be obtained by Ramsey fringes-like experiment, which consists in applying two short, phase coherent, π /2 MW pulses separated by a variable delay time (π/2 refers to π/2 rotation from the pole to the equator of the Bloch sphere). That way, short $T_2$*≈60 ns (corresponding to the free evolution of the spin on the equator of the Bloch sphere) have been measured [Maurand2016]. Dephasing the two short MW pulse can also be used to show that the hole spin can be rotated around two-perpendicular axis on the Bloch sphere. This can be also done by changing the RF frequency between the two pulses. If the source of dephasing fluctuates slowly on the timescale of the hole spin dynamics, spin echo techniques can extend spin coherence: a Hahn echo experiment, where a π pulse around the north pole is introduced half way between the two π /2 pulses, permits to refocus the spin on the Bloch sphere, if the defocusing source is kept constant during the sequence. The fit of the amplitude of the oscillations decay versus the time between the two π /2 pulses gives a coherence time $T_{echo}$ ≈245±12 ns [Maurand2016].

$T_2$* and $T_{echo}$ are relatively short (but much larger than the inverse Rabi frequency) and at least two mechanisms could be invoked as an explanation: the hyperfine coupling with nuclear spin [Koppens05] (Overhauser noise) coming from $^{29}$Si atoms, paramagnetic impurities, boron dopants in the channel, etc. and the charge noise. The distinction between the various mechanisms deserves further studies but it is likely that the charge noise is dominating above the magnetic noise. In particular the gate is very well coupled to the qubit ( a figure-of-merit for the CMOS technology), hence any kind of gate voltage noise is potentially affecting the qubit decoherence time. The sensitivity of the spin to electric noise source is a penalty to pay for using fast electrical manipulation of the hole spin.

The mechanism for EDSR is also not fully clarified in our hole spin quantum bit. It could be g-tensor modulation as explained in the preceding section 3 [Kato03]. It could also involve more standard Rashba spin-orbit couplings [Golovach06]. The magnetic-field angle dependence of spin blockade, EDSR and Rabi frequency are currently studied to measure the g-tensor anisotropy and clarify the mechanism responsible for the observed EDSR.

## 4. Dispersive RF gate-reflectometry and scalable 1D linear array architecture

The next step towards a functional CMOS qubit is to perform fast, high-fidelity, *single shot* qubit read-out. Up to now we measure only the spin orientation in dot 1 by integrating over ≈1s the source-drain current at a repetition rate of 2 MHz. Moreover each time we measure the spin orientation we destroy the phase and the qubit itself. There are two options for single shot readout of the charge transfer-therefore the single spin readout- between the dot 1 and the dot 2: either we put an external high bandwidth charge detector, for instance a nearby single electron transistor [Veldshorst14] or a quantum point contact [Wu14]; or we measure the change of the quantum capacitance for the dot 1 when an hole can be shared between the two dots (at the S(1,1)/S(0,2) degeneracy). The quantum capacitance is directly proportional to the density of states $Cq \propto e^2\, g(E_F)$ which becomes larger at the degeneracy point [Gonzalez16]. This quantum capacitance adds a contribution to the geometrical gate capacitance of dot 1 and dot 2. By detecting the small change in the gate capacitance of dot 2 it is possible to detect if the double dot system is in the S(1,1) or T(1,1) state [Gonzalez16] . We already noticed that the FDSOI trigate technology results in a perfect control of the electrostatic potential in the silicon channel by the gate, that means gate electrode is very well connected to the qubit. Therefore our devices are well adapted to this type of gate capacitance measurements [Gonzalez16] [Crippa17]. The gate capacitance can be measured with a large bandwidth thanks to the so-called dispersive RF reflectometry technique [Ciccarelli11] [Colless13].

A typical RF reflectometry setup and measurement are shown on Fig. 7. Fig. 7a reports a schematic of the dual-port reflectometry circuit, where the two gates of the devices represent two independent readout channels; either circuit comprises a resonator, to maximize the RF signal delivered to the sample, and electronics for the amplification of the signal back reflected. The latter is the demodulated to baseband signal, so that its variations in phase and amplitude point out the modifications in the device admittance. The potentiality of such a technique is shown by Fig. 7b: the phase signal recorded appears also when no source-drain carrier transport takes place, for instance if the tunnelling rates are too small. This property of gate dispersive readout allows to get rid of multiple reservoirs in few-qubit architecture), thereby leading to a tighter qubit pitch as in fig. 8. Fig. 7b also demonstrates that a combination of traces simultaneously acquired from different gate sensors permits to reconstruct the full honeycomb structure of a double quantum dot system where to perform qubit operations.

Nevertheless single shot readout of the spin qubit has not been yet obtained using dispersive RF reflectometry technique. This is currently under study in Grenoble. It will permit to measure the spin qubit on dot 1 using the gate reflectometry on dot 2 for instance, without the need for the drain reservoir (the source reservoir provide holes during the initialization step). This method permits to envision a scalable qubit 1D linear array using the face-to-face arrangement of corner dots along a nanowire[Voisin15] [deFranceschi16], as shown in fig. 8.

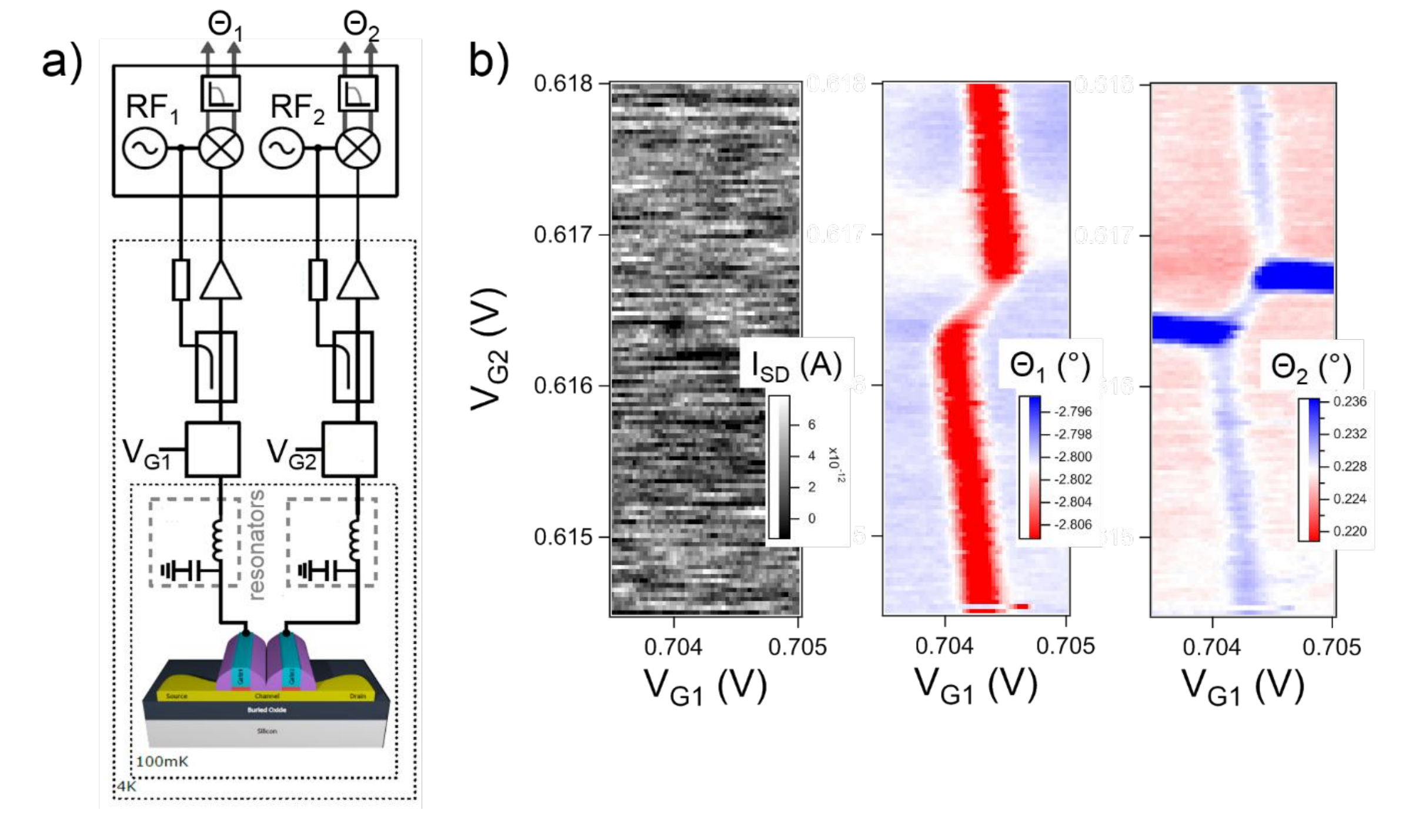

Figure 7: a) Schematic of a dual-port reflectometry setup. Each readout channel is connected to one of the two top gates of the device, so that two independent charge transfer detectors are realized. b) Either of the sensors monitors the single charge transitions involving the dot underneath, though no net source-drain current flows through the transistor; $\Theta_1$ and $\Theta_2$ represent the phase of the signal back reflected by gate 1 and gate 2, respectively.

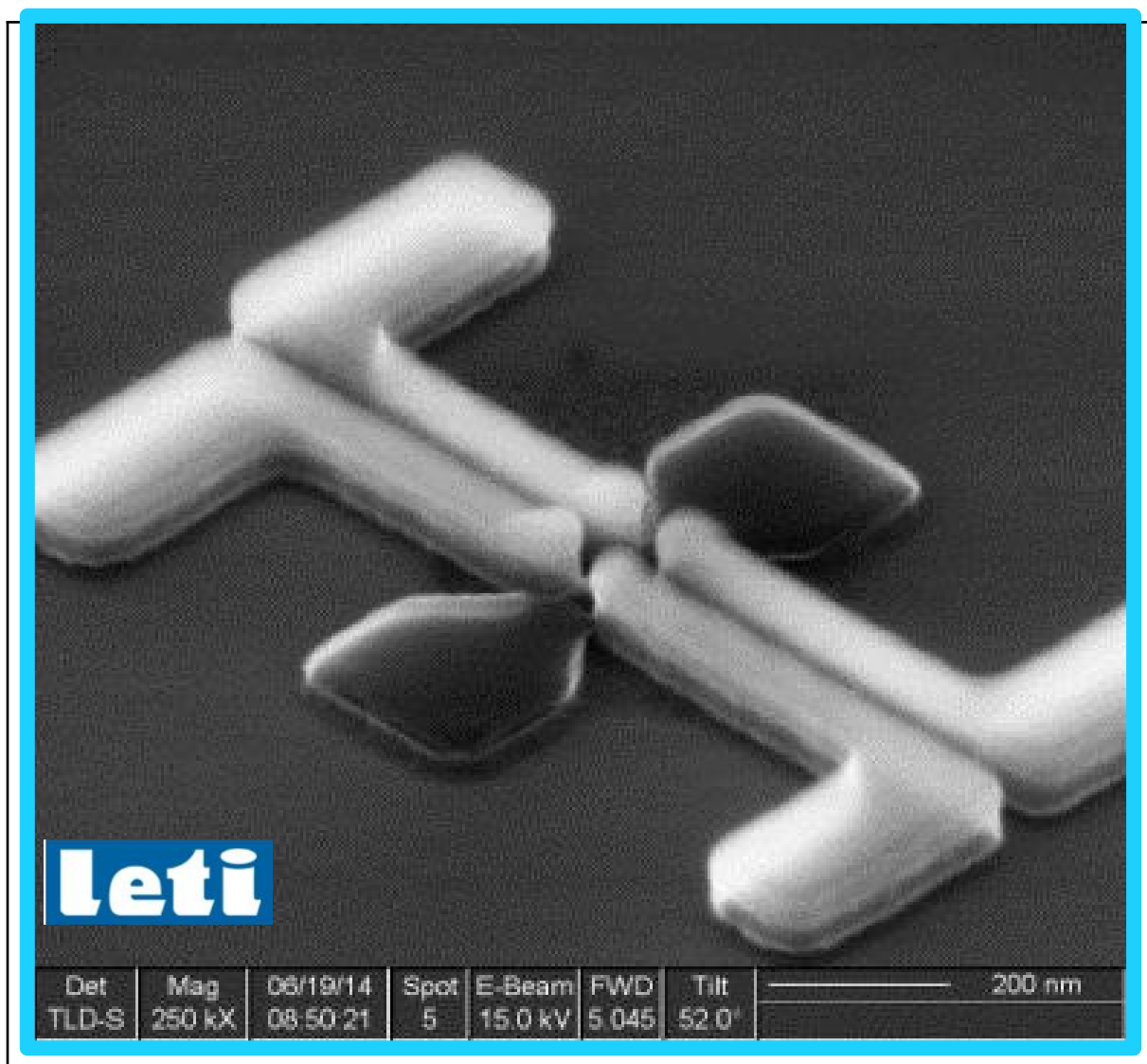

Figure 8: a linear array of CMOS spin qubits. Right panel: the FDSOI nanowire (in blue) is covered with couples of split gates (in grey), forming coupled corner quantum dots indicated by dotted circles [Voisin15]. The raw of A's quantum dot can form the qubits as the raw of B'S quantum dots are used to detect Pauli blockade by dispersive RF gate reflectometry [deFranceschi16]. The A qubits can exchange their spin by nearest neighbour exchange interaction, modulated by the back gate voltage applied across the buried oxide to the substrate. Top panel: A SEM view of a FDSOI nanowire covered by two couples of split gates, fabricated in Grenoble at CEA-LETI.

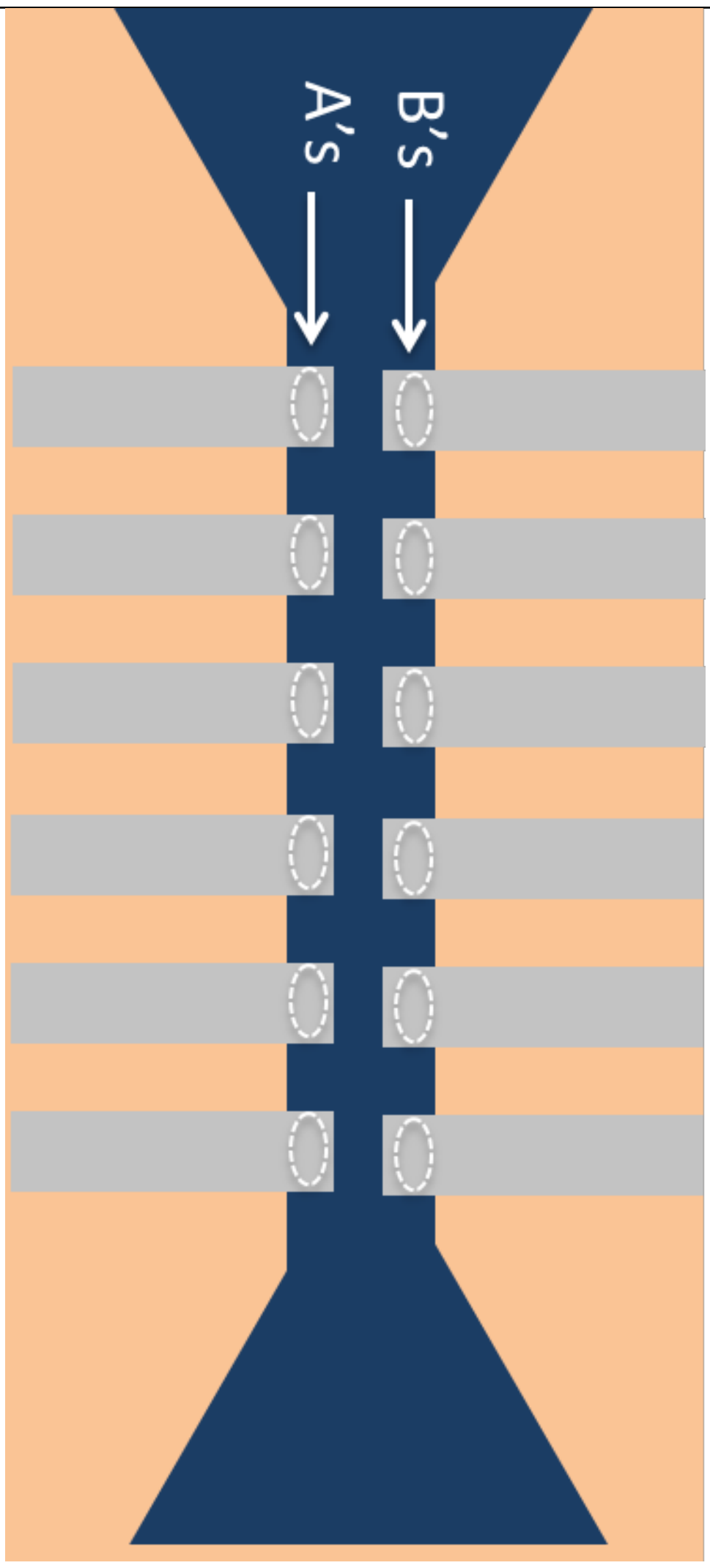

Figure 9 shows the stability diagram for two face-to-face electron corner dots as in fig. 8 recorded at T=4.2K. The stability diagram is recorded at Vb=+30V for which value the two corner dots are relatively strongly coupled. The honeycomb lattice typical for two interacting quantum dots is clearly visible. The two dots are in parallel such that a drain-source current is measured along the line where the number of electrons is degenerate on one of the two dots (in contrast for two dots in series with negligible co-tunneling effect the current will appear only at the intersections of the lines, the so-called triple points, evolving to triangles at finite Vds as in fig. 5). For electrons it is possible to know the exact occupation number of the corner dots, which constitute therefore a clean platform for implementing an electron spin CMOS qubit and a 1D array as featured in fig. 8.

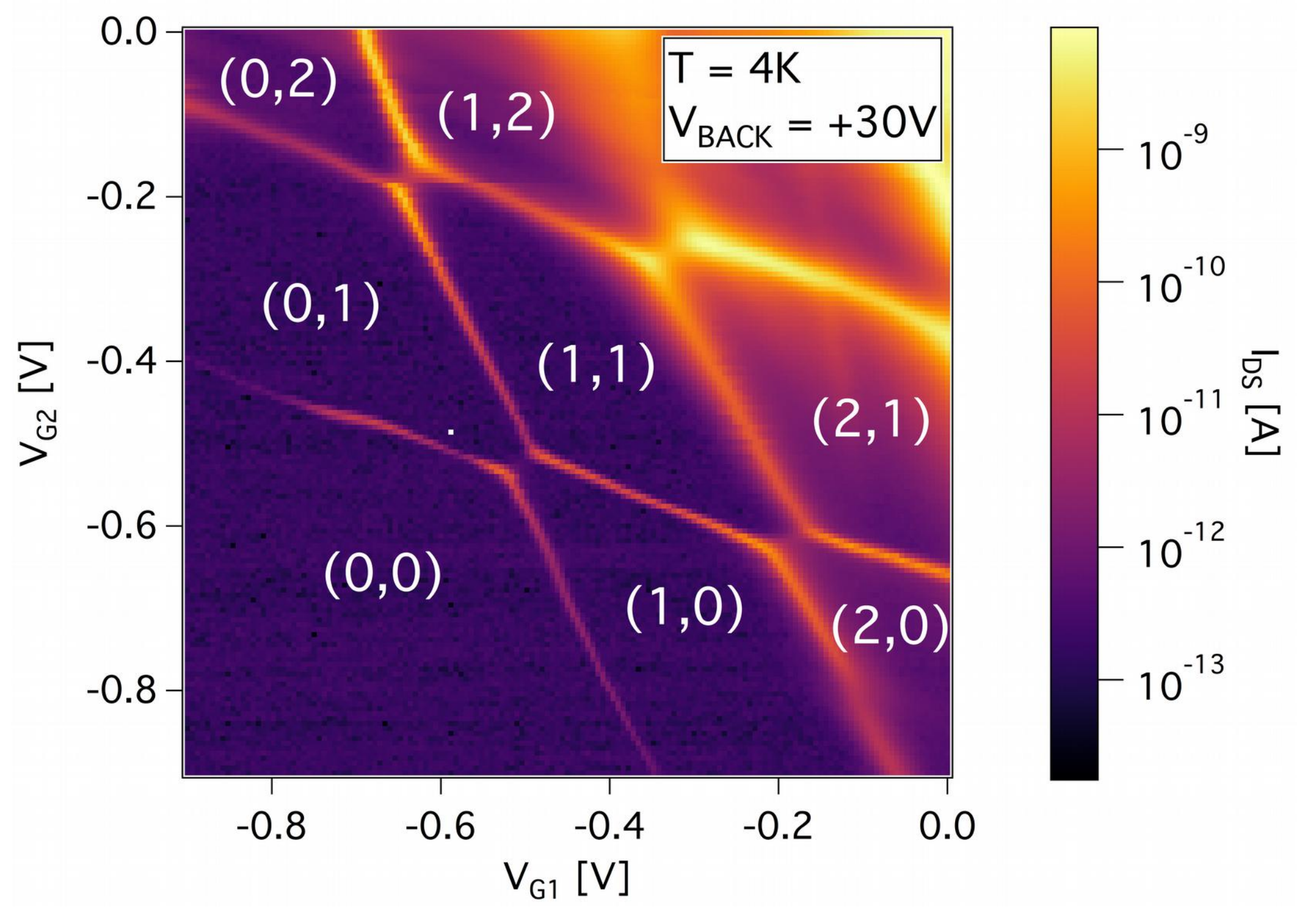

Figure 9: top: Stability diagram for two face-to-face corner dots: the drain source current is recorded as function of the two gate voltages applied on the split gate (T=4.2K , $V_{DS}$ = 3mV and $V_{back}$ (substrate bias) = +30V). The current is zero when the electron number on both dot is fixed. The current lines delimitate regions where the number (n,m) of electrons on both dots is constant. right: sketch of the cross-section of the silicon nanowire (yellow) covered by the split gate (red). The two coupled quantum dots are featured as violet circles. From

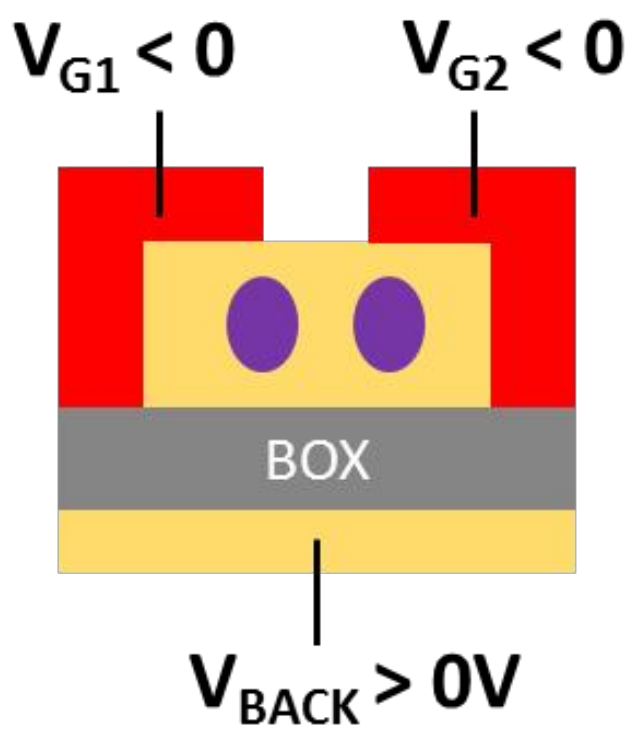

| | |
|---|---|
| [deFranceschi16] | |

## 5. Summary

By adapting the most advanced CMOS technology –available only in pre-industrial and industrial platforms- we have been able to make a CMOS device working as a spin qubit at low temperature. The spin qubit is based on PMOS type of nanowire field effect transistor that permits for the first time to realize an hole spin qubit. The decisive advantage for using holes is that the spin can be manipulated by electric voltage applied on standard gate, thanks to spin-orbit coupling. This breakthrough provides an entirely new way to envision a CMOS quantum computer core which is fully co-integrable with classical CMOS cryo-electronic peripherals. The realization of a scalable multi-qubit layout remains to be demonstrated. One important, possibly indispensable ingredient is a means to extend the spatial distance over which qubits can be coupled. Another topic that requires continued attention for all semiconductor qubits is the realization of high fidelity qubit manipulation, particularly for two-qubit gates, and scalable readout.

**Acknowledgements:** The research leading to these results has been supported by the European Community's H2020 Framework under the project MOSQUITO "MOS based quantum information Technology " grant number 688539.

***Reference Citations***